\documentstyle[sprocl,epsfig,color,graphics]{article}

\def\Journal#1#2#3#4{{#1} {\bf #2}, #3 (#4)}

\def\NPB{{\em Nucl. Phys.} B}
\def\PLB{{\em Phys. Lett.}  B}

\def\PRD{{\em Phys. Rev.} D}
\def\ZPC{{\em Z. Phys.} C}

\def\b               {\beta}
\def\g               {\gamma}

\def\s               {\sigma}
\def\t               {\theta}
\def\x               {\chi}
\def\G               {\Gamma}
\def\D               {\Delta}

\def\ti    {\tilde}
\def\sf    {{\tilde f}}
\def\sq    {{\tilde q}}
\def\st    {{\tilde t}}
\def\sb    {{\tilde b}}
\def\stau  {{\tilde\tau}}
\def\snu   {{\tilde\nu}}
\def\sl    {{\tilde \ell}}
\def\ch    {\ti \x}
\def\nt    {\ti \x^0}
\def\sg    {\ti g}

\def\cst   {\cos\theta_\st}
\def\csb   {\cos\theta_\sb}

\newcommand{\mst}[1]   {m_{\ti t_{#1} }}
\newcommand{\msb}[1]   {m_{\ti b_{#1} }}
\newcommand{\mstau}[1] {m_{\ti \tau_{#1} }}
\newcommand{\mch}[1]   {m_{\ti \chi^+_{#1} }}

\def\Pm  {{\cal P}_-}
\def\Pp  {{\cal P}_+}
\def\fbi {{\rm fb}^{-1}}

\newcommand{\gsim}{\;\raisebox{-0.9ex}
           {$\textstyle\stackrel{\textstyle >}{\sim}$}\;}

\newcommand{\lsim}{\;\raisebox{-0.9ex}{$\textstyle\stackrel{\textstyle<}
           {\sim}$}\;}

\begin{document}
\setlength{\unitlength}{1mm}

\title{PRODUCTION AND DECAYS OF STOPS, SBOTTOMS, AND STAUS}

\author{\underline{H. Eberl}, S. Kraml, W. Majerotto}

\address{Inst. f. Hochenergiephysik der \"OAW, \\
  Nikolsdorfer G. 18, A--1050 Vienna, Austria}

\author{A. Bartl, W. Porod}

\address{Inst. f. Theoretische Physik, Univ. Wien, \\
   Boltzmanng. 5, A--1090 Vienna, Austria}

\maketitle\abstracts{
We present a phenomenological analysis of production and
decays of $\st$, $\sb$, and $\stau$ in $e^+ e^-$~collisions with
$\sqrt{s} = 500$ -- 800~GeV. We include SUSY--QCD and Yukawa coupling 
corrections as well as initial state radiation. We show
that $e^-$~beam polarization is a powerful tool for a determination of
the underlying SUSY parameters. Using in addition a polarized
$e^+$~beam improves the precision of the parameter determination by
about 25\%.}
 
%------------------------------------------------------------------------------
\section{Introduction}
%------------------------------------------------------------------------------

Supersymmetry (SUSY) implies the existence of two scalar partners
$\sq_L$, $\sq_R$ (squarks) and $\sl_L$, $\sl_R$ (sleptons) to each
quark $q$ and lepton $\ell$, respectively. The squarks and sleptons of
the third generation are of special interest due to their sizeable
Yukawa couplings being proportional to $m_q$ or $m_\ell$. 
These may induce 
a large mixing between $\sf_L$ and $\sf_R$ (with $\sf = \sq$ or
$\sl$), $\sf_1 = \cos\theta_\sf \sf_L + \sin\theta_\sf \sf_R$,
$\sf_2 = - \sin\theta_\sf \sf_L + \cos\theta_\sf \sf_R$, where
$\sf_{1,2}$ are the mass eigenstates ($m_{\sf_1} < m_{\sf_2}$). In
particular, the stop $\st_1$ is most likely the lightest squark due to
the large top mass. Also $\sb_1$ and $\stau_1$ can be relatively
light for large $\tan\beta = v_2/v_1$ (where $v_1$ and $v_2$ are the
vacuum expectation values of the two Higgs doublets). 
%Moreover, one expects from RG equations that the soft SUSY breaking
%parameters $M_{\tilde Q}$, $M_{\tilde U}$, and $M_{\tilde D}$ of the
%third generation are smaller than those of the first and second generation.
Hence these
particles might be produced in $e^+ e^-$~collisions at the next linear
collider with a center of mass energy $\sqrt{s} \gsim 500$~GeV. 
Their properties could be determined precisely enough to test
supersymmetric models.
In this contribution we want to give a rather complete 
phenomenological analysis
of the production and decays of stops, sbottoms, and staus. In
particular, we discuss the usefulness of polarisation of the $e^-$ and
$e^+$~beams for the determination of the underlying SUSY parameters.

%------------------------------------------------------------------------------
\section{Production cross section}
%------------------------------------------------------------------------------

\begin{figure}[h!]
\begin{center}
\begin{picture}(52,30)
%\put(0,0){\framebox(52,30)}
\put(4,0){\mbox{\epsfig{figure=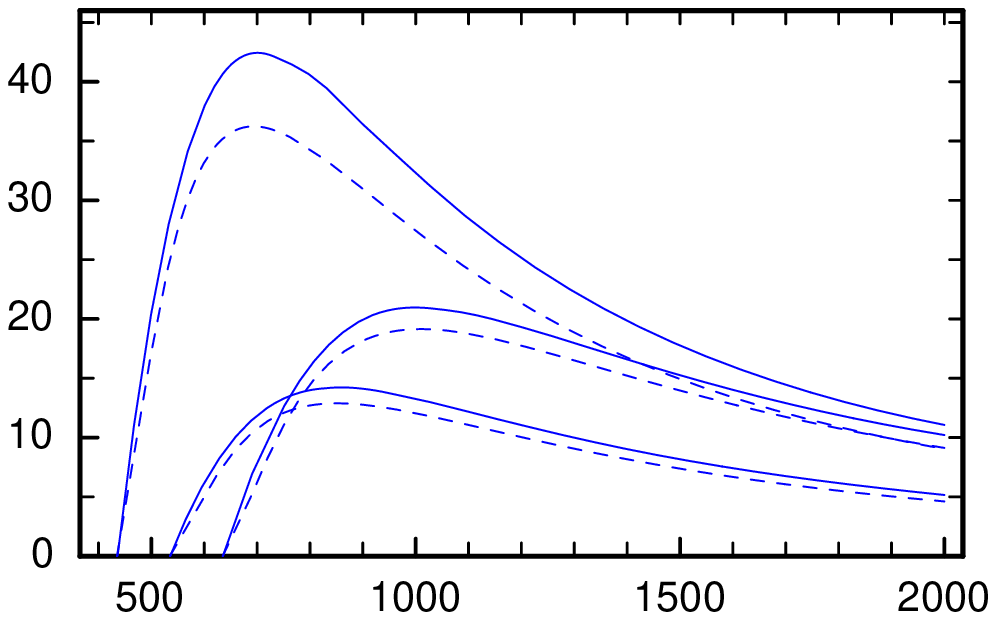,height=30mm}}}
\put(-1,5){\rotatebox{90}{\scriptsize $\s(e^+e^-\to\st_i\bar{\st_j}$)~[fb]}}
\put(25,-3){\mbox{\scriptsize $\sqrt{s}$ [GeV]}}
\put(24,24){\mbox{\scriptsize $\st_1\bar{\st_1}$}}
\put(16.5,16){\mbox{\scriptsize $\st_2\bar{\st_2}$}}
\put(24.5,6){\mbox{\scriptsize $\st_1\bar{\st_2}$ + c.c.}}
\end{picture}\hspace{5mm}
\begin{picture}(52,30)
%\put(0,0){\framebox(52,30)}
\put(4,0){\mbox{\epsfig{figure=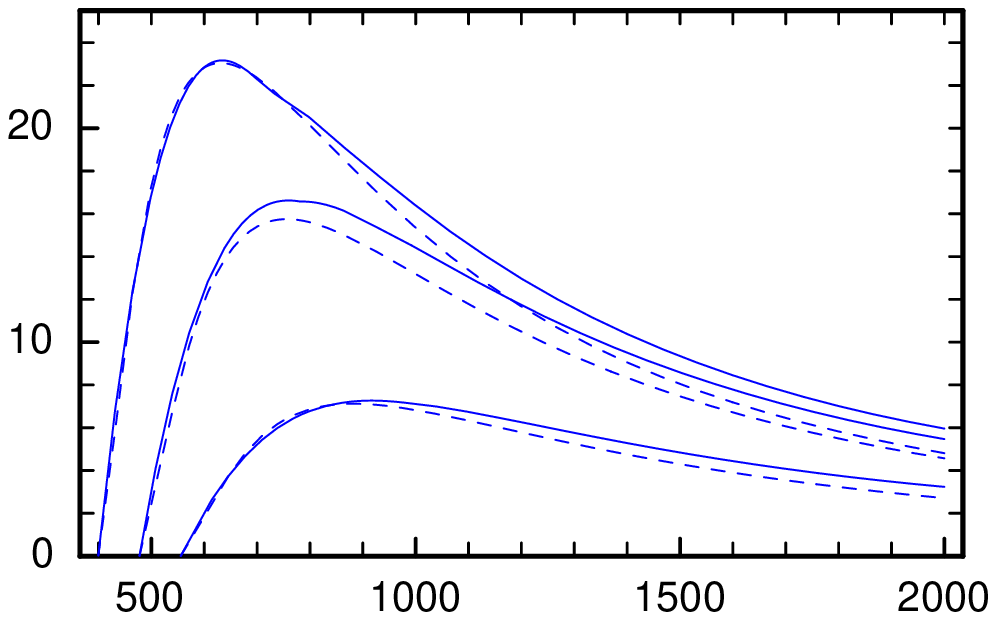,height=30mm}}}
\put(-1,5){\rotatebox{90}{\scriptsize $\s(e^+e^-\to\sb_i\bar\sb_j$)~[fb]}}
\put(25,-3){\mbox{\scriptsize $\sqrt{s}$ [GeV]}}
\put(22.5,24){\mbox{\scriptsize $\sb_1\bar\sb_1$}}
\put(18,7){\mbox{\scriptsize $\sb_2\bar\sb_2$}}
\put(15,14){\mbox{\scriptsize $\sb_1\bar\sb_2$ + c.c.}}
\end{picture}\hspace{5mm}
\end{center}
\caption{Cross sections for $e^+e^-\to\st_i\bar\st_j$ and 
$e^+e^-\to\sb_i\bar\sb_j$ as a function of $\sqrt{s}$ 
for $\mst{1}=218$ GeV, $\mst{2}=317$ GeV, $\cst=-0.64$, 
$\msb{1}=200$ GeV, $\msb{2}=278$ GeV, $\csb=0.79$, 
$M=200$ GeV, $\mu=1000$ GeV, $\tan\b=4$, and $m_A=300$ GeV; 
the dashed lines show the tree level and the full lines 
the total corrected cross sections.}
\label{fig:Xsect}
\end{figure}

\begin{figure}[h!]
\begin{center}
\begin{picture}(50,30)
%\put(0,0){\framebox(50,30)}
\put(-3,-6){\mbox{\epsfig{figure=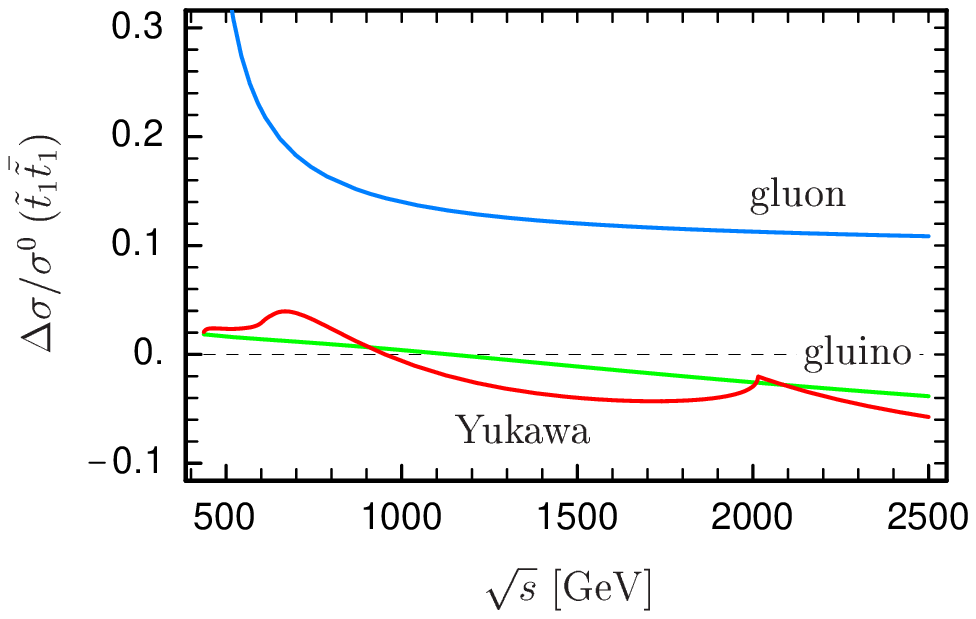,height=40mm}}}
\end{picture}\hspace{5mm}
\begin{picture}(50,30)
%\put(0,0){\framebox(50,30)}
\put(-3,-6){\mbox{\epsfig{figure=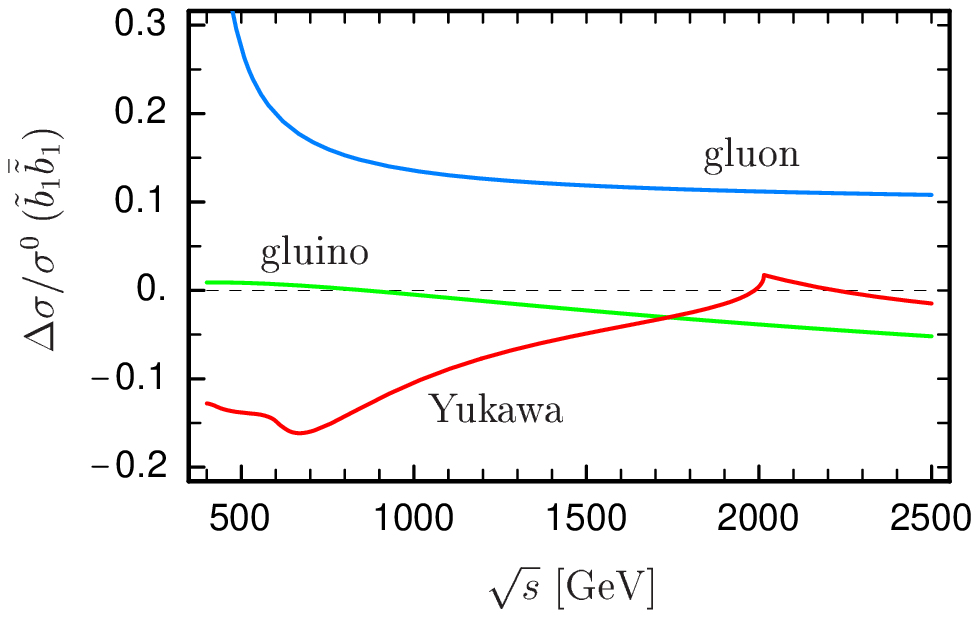,height=40mm}}}
\end{picture}
\end{center}
\caption{Gluon, gluino, and Yukawa coupling corrections~$^4$
%~\cite{ekm} 
for $e^+e^-\to\st_1\bar\st_1$ and $e^+e^-\to\sb_1\bar\sb_1$
relative to the tree--level cross section 
for the parameters of  Fig.~\ref{fig:Xsect}.}
\label{fig:relcorr}
\end{figure}

At an $e^+e^-$ collider, $\st$, $\sb$, and $\stau$ can be pair--produced 
via $\g$ and $Z$ exchange in the s--channel. 
The total cross section shows the typical $\b^3$ dependence, where 
$\b$ is the velocity of the outgoing sfermions. 
It has turned out that conventional QCD corrections~\cite{hk,dh} as
well as SUSY--QCD corrections \cite{ebm} and 
Yukawa coupling corrections \cite{ekm} are important. Moreover, it is
necessary to take into account initial state radiation~\cite{dh} (ISR).
As an example, we compare in Fig.~1 the $\sqrt{s}$ dependence of the
tree level cross sections of $e^+ e^-\to\st_i\bar\st_j$ and 
$e^+ e^-\to\sb_i\bar\sb_j$ with the total (SUSY--QCD, Yukawa coupling,
and ISR) corrected cross sections 
for $\mst{1}=218$ GeV, $\mst{2}=317$ GeV, $\cst=-0.64$, 
$\msb{1}=200$ GeV, $\msb{2}=278$ GeV, $\csb=0.79$, 
$M=200$ GeV, $\mu=1000$ GeV, $\tan\b=4$, and $m_A=300$ GeV. 
Figure~\ref{fig:relcorr} shows the gluon, gluino, 
and Yukawa coupling corrections for $\st_1\bar{\st_1}$ 
and $\sb_1\bar\sb_1$ production relative to the tree level cross 
section for the parameters of  Fig.~\ref{fig:Xsect}. 
As can be seen, all three contributions can be significant 
for precision measurements. 
In the case of stau production, Yukawa coupling corrections are   
typically $\lsim 5\%$. 
ISR changes the cross section by up to $\sim 25\%$. 

Beam polarization can be used to enhance the signal and reduce the background.
Figure~\ref{fig:pol-err}\,a shows $\s(e^+e^-\to\st_1\bar{\st_1})$ in the 
$\Pm$--$\,\Pp$ plane, with $\Pm$ the polarization of the $e^-$ beam and 
$\Pp$ that of $e^+$ beam
(${\cal P}_\pm = \{-1,0,1\}$ for \{left--, un--, right--\} polarized),
for $\mst{1}=200$ GeV, $\cst=-0.66$, 
and $\sqrt{s}=500$ GeV. 
For the SUSY--QCD and Yukawa coupling corrections we have used 
$\mst{2}=420$ GeV, $\msb{1}=297$ GeV, $\msb{2}=345$ GeV, $\csb=0.84$, 
$M=200$ GeV, $\mu=800$ GeV, $m_A=300$ GeV, and $\tan\b=4$. 
ISR has also been taken into account. 
The non--shaded area is the range of polarization of the TESLA  
design \cite{tesla}. 

We next estimate the precision one may obtain 
for the stop parameters from cross section measurements. 
We use the parameter point of Fig.~\ref{fig:pol-err}\,a, 
i.e. $\mst{1}=200$ GeV, $\mst{2}=420$ GeV, 
$\cst=-0.66$, etc. as an illustrative example: 
For 90\% left--polarized electrons (and unpolarized positrons) we have  
$\s_L(\st_1\bar\st_1)=44.88$\,fb. 
For 90\% right--polarized electrons we have $\s_R(\st_1\bar\st_1)=26.95$\,fb.
According to the Monte Carlo study of \cite{nowak} 
one can expect to measure these cross sections 
with an error of $\D\s_L = \pm 2.1\,\%$ and $\D\s_R = \pm 2.8\,\%$ 
in case of an integrated luminosity of ${\cal L}=500\,\fbi$ 
(i.e. ${\cal L}=250\,\fbi$ for each polarization).
Scaling these values to ${\cal L}=100\,\fbi$ leads to 
$\D\s_L = \pm 4.7\,\%$ and $\D\s_R = \pm 6.3\,\%$.  
Figure~\ref{fig:pol-err}\,b shows the corresponding error bands and 
error ellipses in the $\mst{1}$--$\,\cst$ plane. 
The resulting errors on the stop mass and mixing angle are:  
$|\D\mst{1}|=2.2$ GeV, $|\D\cst|= 0.02$ for ${\cal L}=100\,\fbi$ 
and $|\D\mst{1}|=0.98$ GeV, $|\D\cst|= 0.01$ for ${\cal L}=500\,\fbi$. 
With the additional use of a 60\% polarized $e^+$ beam these values can still 
be improved by $\sim 25\%$. 
At $\sqrt{s}=800$ GeV also $\st_2$ can be produced: 
$\s(\st_1\bar\st_2+\mbox{c.c.})=8.75$\,fb for $\Pm=-0.9$ and $\Pp=0$. 
If this cross section can be measured with a precision of $\pm 6\%$   
this leads to $\mst{2}=420\pm 8.3$ GeV. 
\footnote{Here note that $\st_1\bar\st_1$ is produced at $\sqrt{s}=800$ GeV with 
an even higher rate than at $\sqrt{s}=500$ GeV. 
One can thus improve the errors on $\mst{1}$, $\mst{2}$, and $\cst$ 
by combining the information obtained at different energies. 
However, we will not do this in this study.}
If $\tan\b$ and $\mu$ are known from other measurements 
this then allows one to 
determine the soft SUSY breaking parameters of the stop sector.  
Assuming $\tan\b=4\pm 0.4$ leads to $M_{\ti Q}=298.2\pm 7.3$ GeV and 
$M_{\ti U}=264.4\pm 6.7$ GeV. 
In addition, assuming $\mu=800\pm 80$ GeV we get $A_t=586.5\pm 34.5$ 
(or $-186.5 \pm 34.5$) GeV.
%with $\D\mu\to 0$ $|\D A_t|\to 25.6$ GeV.
The ambiguity in $A_t$ exists because the sign of
$\cst$ can hardly be determined from cross section measurements. 
This may, however, be possible from measuring decay branching ratios 
or the stop--Higgs couplings. 

\fboxsep 2pt
\definecolor{gray}{rgb}{0.7,0.7,0.7}
\begin{figure}
\begin{center}
\begin{picture}(45,45)
%\put(0,0){\framebox(45,45)}
\put(0,0){\mbox{\epsfig{figure=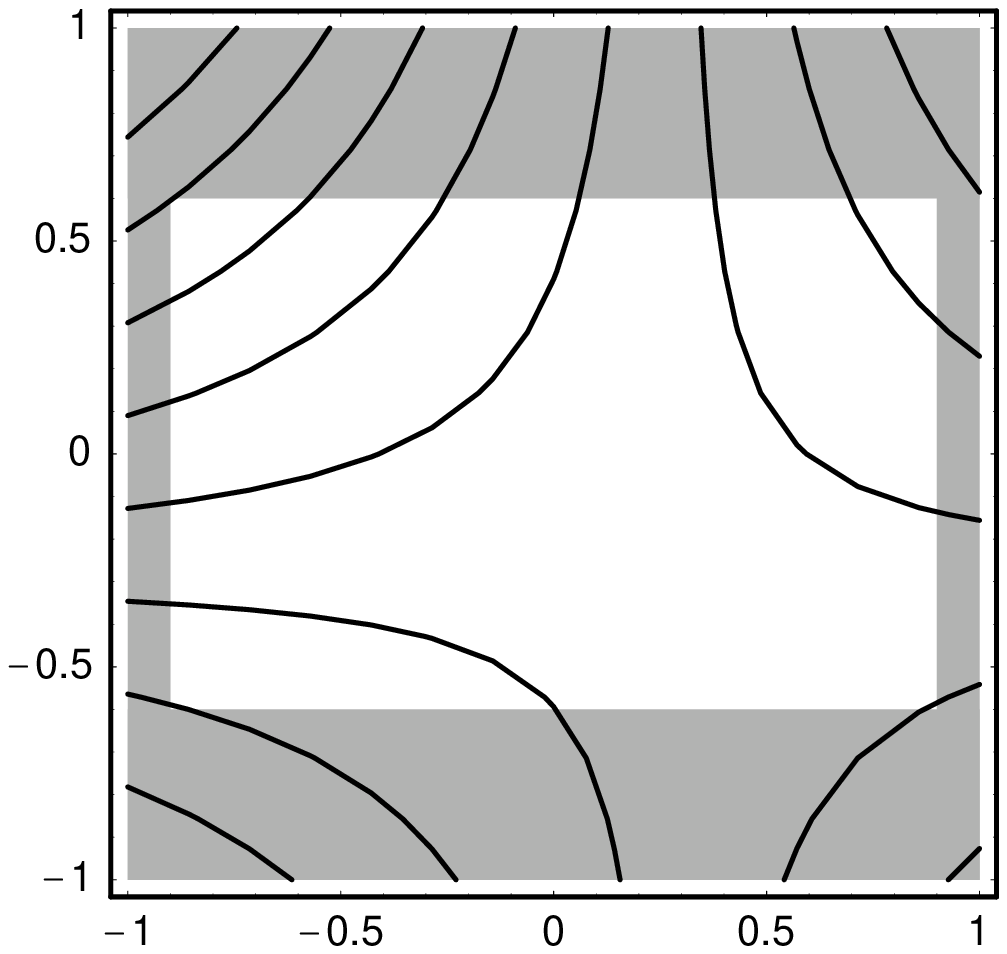,height=45mm}}}
\put(-4,40){\mbox{(a)}}
\put(23.5,-2){\mbox{\footnotesize $\Pm$}}
\put(-4,23){\rotatebox{90}{\footnotesize $\Pp$}}
\put(17,24){\colorbox{white}{\scriptsize 40}}
\put(12,28){\colorbox{white}{\scriptsize 50}}
\put(9,32){\colorbox{white}{\scriptsize 60}}
\put(8,36){\fboxsep 1pt\colorbox{gray}{\scriptsize 70}}
\put(7,40){\fboxsep 1pt\colorbox{gray}{\scriptsize 80}}
\put(17,15){\colorbox{white}{\scriptsize 30}}
\put(13,9){\colorbox{gray}{\scriptsize 20}}
\put(8,6){\colorbox{gray}{\scriptsize 10}}
\put(35.5,8){\colorbox{gray}{\scriptsize 40}}
\put(33,24){\colorbox{white}{\scriptsize 30}}
\put(37,32){\colorbox{white}{\scriptsize 20}}
\put(39.5,39){\colorbox{gray}{\scriptsize 10}}
\end{picture}\hspace{12mm}
\begin{picture}(45,45)
%\put(0,0){\framebox(45,45)}
\put(0,0){\mbox{\epsfig{figure=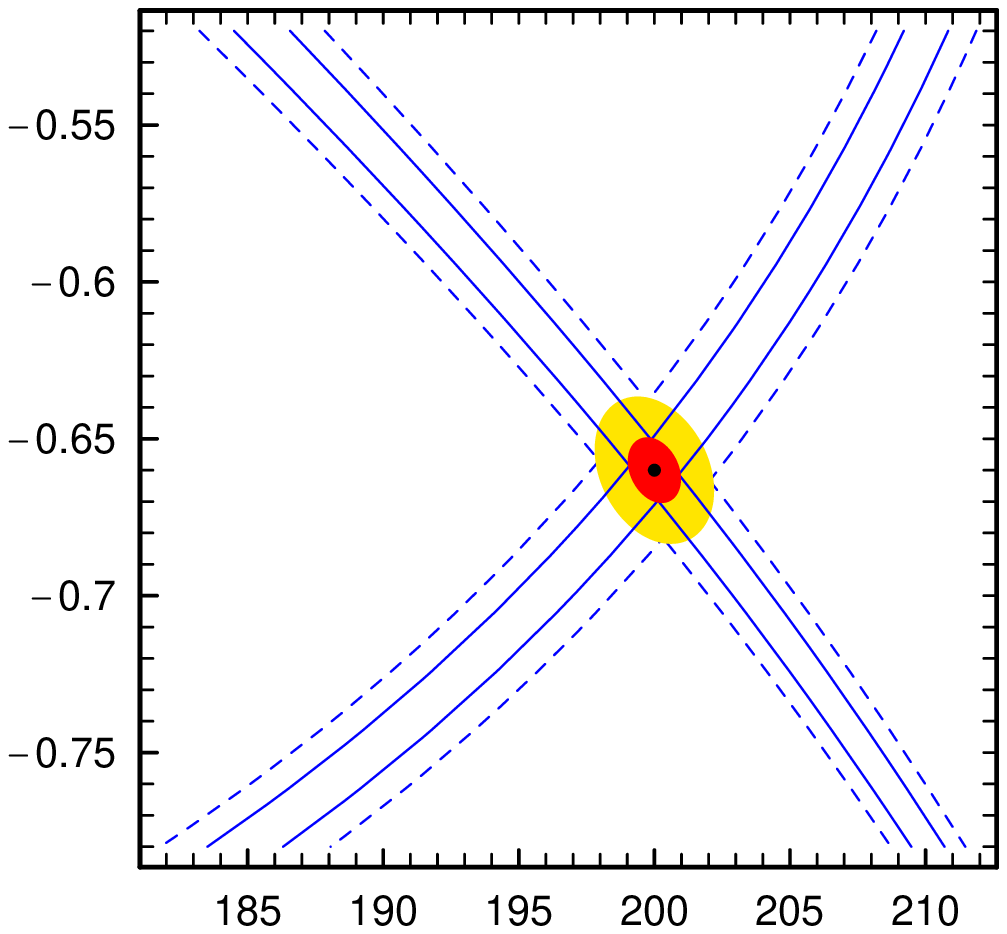,height=45mm}}}
\put(-4,40){\mbox{(b)}}
\put(19,-2){\mbox{\scriptsize $\mst{1}$ [GeV]}}
\put(-4,21){\rotatebox{90}{\scriptsize $\cst$}}
\put(10,36){\colorbox{white}{\scriptsize $\Pm=-0.9$}}
\put(10,9){\colorbox{white}{\scriptsize $\Pm=+0.9$}}
\end{picture}
\end{center}
\caption{(a) Dependence of $\s(e^+e^-\to\st_1\bar{\st_1})$ (in fb) on the beam 
polarization, for $\mst{1}=200$ GeV, $\cst=-0.66$, $\sqrt{s}=500$ GeV, 
and the other parameters as given in the text.
(b) Error bands and 68\% CL error ellipse for determining $\mst{1}$ and $\cst$ 
from cross section measurements, for $\Pm=\pm 0.9$, $\Pp=0$, 
and the other parameters as in (a); the dashed lines are for 
${\cal L}=100\,\fbi$ and the full lines for ${\cal L}=500\,\fbi$.}
\label{fig:pol-err}
\end{figure}
  
A method to reconstruct the squark mass by the decay kinematics has
been proposed in \cite{feng}. 
Though their analysis was performed for squarks of the 1st
and 2nd generation it is also applicable to the 3rd generation, and a
precision similar to the one presented here can be expected. 
A Monte Carlo study of stau production, with $\stau_1\to\tau\nt_1$, 
was performed in \cite{nojiri}. They also give a method for 
the parameter determination concluding that $m_{\stau_1}$ and $\t_{\stau}$ 
could be measured within an error of few percent.

%------------------------------------------------------------------------------
\section{Decays} 
%------------------------------------------------------------------------------

\begin{figure}[h!]
\begin{center}
\begin{picture}(56,34)
%\put(0,0){\framebox(56,34)}
\put(4,0){\mbox{\epsfig{figure=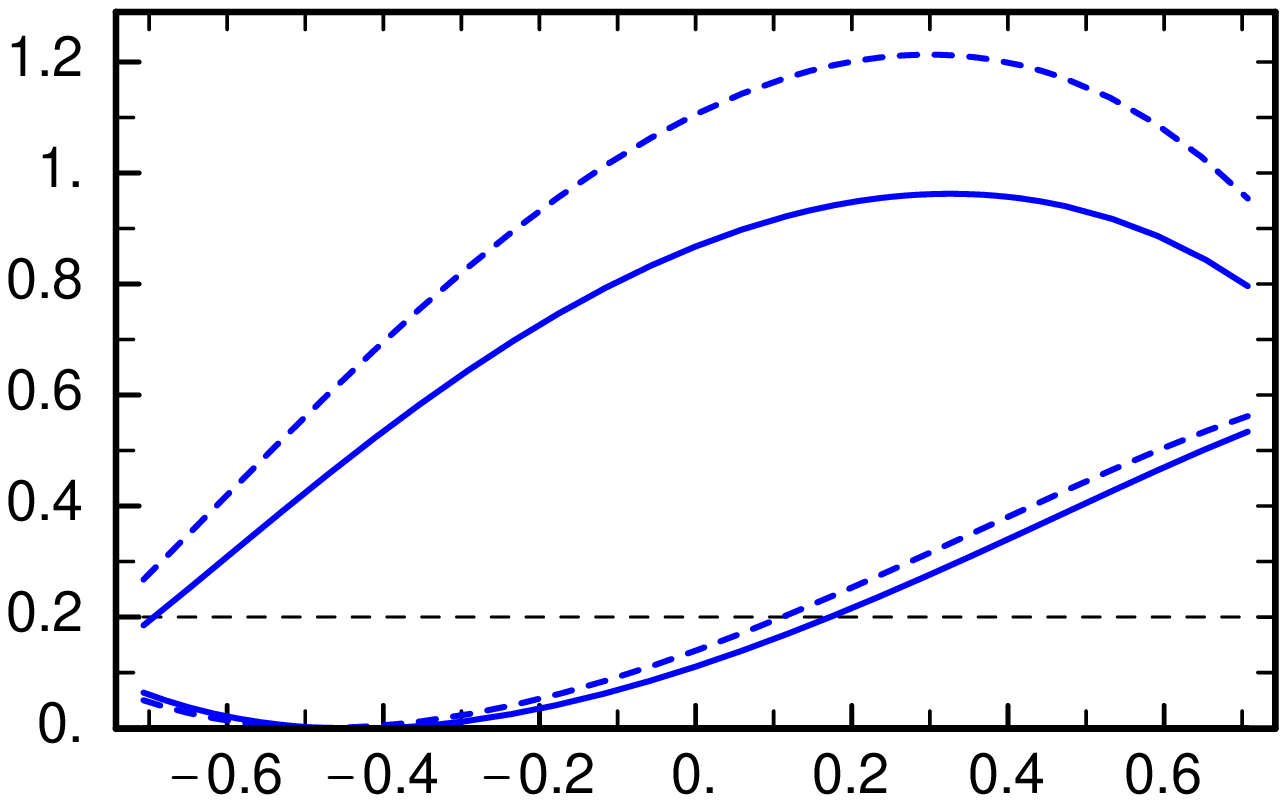,height=35mm}}}
\put(-1,8){\rotatebox{90}{\scriptsize $\G(\st_1\to b\ch^+_1$)~[GeV]}}
\put(28,-2.5){\mbox{\scriptsize $\cst$}}
\put(37,13){\mbox{\scriptsize (a)}}
\put(35,27){\mbox{\scriptsize (b)}}
\end{picture}
\end{center}
\caption{Decay width of $\st_1$ as a function of $\cst$ for $\mst{1}=200$ GeV, 
$\mst{2}=490$ GeV, $\mch{1}=133$ GeV and $\tan\b=3$; 
in (a) $M=160$ GeV, $\mu=300$ GeV and in (b) $M=300$ GeV, $\mu=160$ GeV. 
The dashed lines are the tree--level 
and the full lines the SUSY--QCD corrected results.}
\label{fig:Gstop1}
\end{figure}

\begin{figure}[h!]
\begin{center}
\begin{picture}(120,34)
%\put(0,0){\framebox(120,34)}
\put(0,-4){\mbox{\epsfig{figure=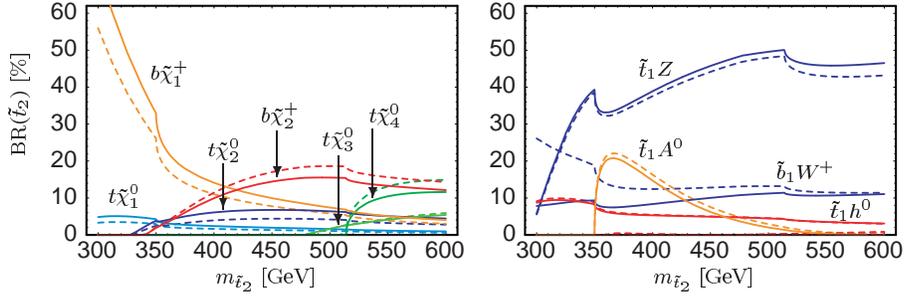,height=41mm}}}
\end{picture}
\end{center}
\caption{Branching ratios of $\st_2$ decays as a function of $\mst{2}$ 
for $\mst{1}=200$ GeV, $\cst=0.6$, $M=180$ GeV,  $m_A=150$ GeV, $\tan\b=3$,
$\mu=300$ GeV, $A_t=A_b$, and $M_{\ti D}=1.1\,M_{\ti Q}$; 
the dashed lines show the tree level and the full lines the SUSY--QCD 
corrected results.}
\label{fig:BRstop2}
\end{figure}

\begin{figure}[h!]
\begin{center}
\begin{picture}(50,45)
%\put(0,0){\framebox(50,45)}
\put(0,-4){\mbox{\epsfig{figure=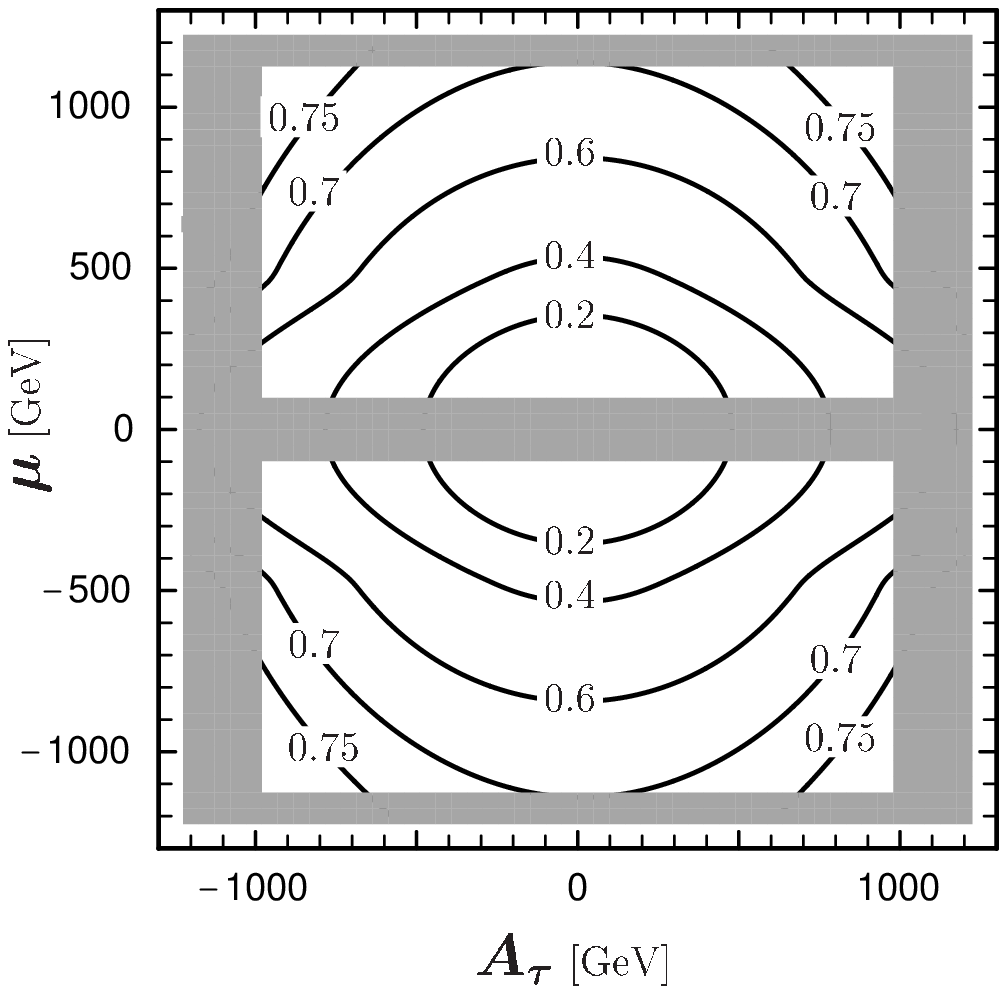,height=54mm}}}
\put(-4,41){\mbox{(a)}}
\end{picture}\hspace{10mm}
\begin{picture}(50,45)
%\put(0,0){\framebox(50,45)}
\put(-8,-38){\mbox{\epsfig{figure=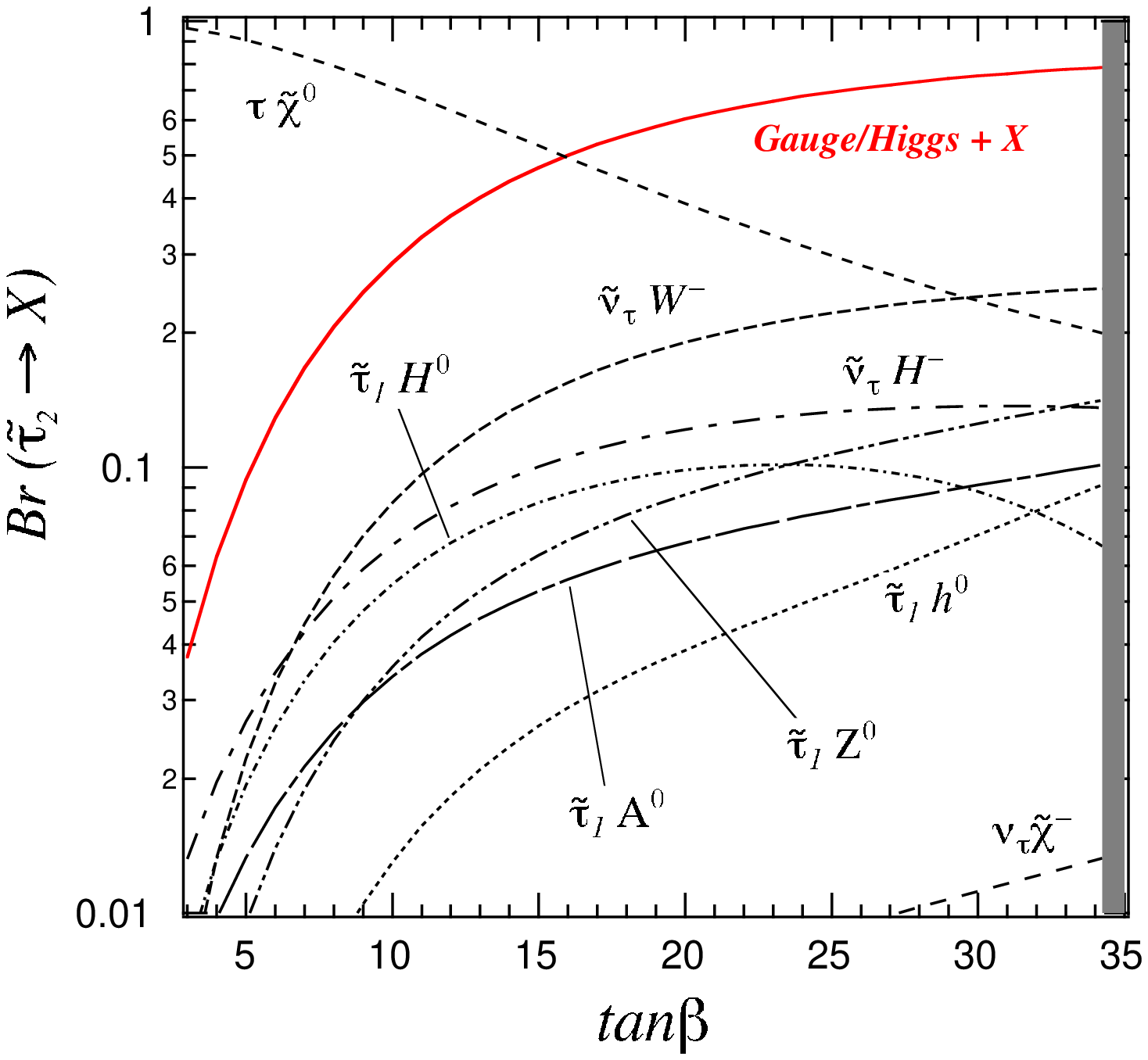,height=100mm}}}
\put(-3,41){\mbox{(b)}}
\end{picture}
\end{center}
\caption{Branching ratios of $\stau_2$ decays for $\mstau{1}=250$ GeV, 
$\mstau{2}=500$ GeV, and $\mstau{L}<\mstau{R}$: 
(a) $\sum \mbox{BR}[\stau_2\to \stau_1 + (Z^0,\,h^0,\,H^0,\,A^0), 
                             \:\snu_\tau + (W^-,\,H^-)]$ for $\tan\b = 30$;
(b) $\tan\b$ dependence of the individual branching ratios 
for $A_\tau=800$ GeV and $\mu=1000$ GeV; 
the other parameters are as in Fig.~\ref{fig:BRstop2}.}
\label{fig:BRstau}
\end{figure}

Stop, sbottoms, and staus can have quite complicated decay modes 
\cite{bmp,sq2dec,staudec}. 
In addition to the decays into fermions, 
i.e. into chargino, neutralino or gluino:
\begin{equation}
\begin{array}{lll}
  \st_i\to b\ch^+_j, \qquad   & \st_i\to t\nt_k, \qquad & \st_i\to t\sg, \\
  \sb_i\to t\ch^-_j,          & \sb_i\to b\nt_k,        & \sb_i\to b\sg, \\
  \stau_i\to \nu_\tau\ch^-_j, &  \stau_i\to \tau\nt_k,  & 
\end{array}
\label{eq:fermdec}
\end{equation}
$(i,j=1,2;\,k=1...4)$ they may also decay into bosons, 
i.e. into a lighter sfermion plus a gauge or Higgs boson: 
\begin{equation}
\begin{array}{ll}
  \st_i\to \sb_j + (W^+,\,H^+),\qquad 
    & \st_2\to \st_1 + (Z^0,\,h^0,\,H^0,\,A^0),\\
  \sb_i\to \st_j + (W^-,\,H^-), 
    & \sb_2\to \sb_1 + (Z^0,\,h^0,\,H^0,\,A^0),\\
  \stau_i\to \snu_\tau + (W^-,\,H^-), 
    & \stau_2\to \stau_1 + (Z^0,\,h^0,\,H^0,\,A^0),
\end{array}
\label{eq:bosdec}
\end{equation}
provided the mass splitting is large enough. 
If the 2--body decays (\ref{eq:fermdec}) and (\ref{eq:bosdec}) 
are kinematically 
forbidden, loop decays~\cite{hk}, 3--body~\cite{hodec3}, 
and even 4--body~\cite{hodec4} decays come into play. 
For $\st$ and $\sb$ decays SUSY--QCD corrections can be important 
\cite{qcdsg,qcdnc,qcdwz,qcdhx}. 
In Fig.~\ref{fig:Gstop1} we plot the decay width of $\st_1\to b\ch^+_1$ 
as a function of $\cst$ for $\mst{1}=200$ GeV, 
$\mst{2}=490$ GeV, $\mch{1}=133$ GeV and $\tan\b=3$. 
Two cases are shown: one for $\ch^+_1\sim \tilde W^+$ ($M\ll|\mu|$) 
and one for $\ch^+_1\sim \tilde H^+$ ($M\gg|\mu|$). 
In this figure the decay $\st_1\to b\ch^+_1$ has 
practically 100\% branching ratio 
except where the tree--level coupling vanishes. 
Figure~\ref{fig:BRstop2} shows the decay branching ratios of $\st_2$ 
decays as a function of $\mst{2}$ 
for $\mst{1}=200$ GeV, $\cst=0.6$, $M=180$ GeV,  $m_A=150$ GeV, $\tan\b=3$,
$\mu=300$, $A_t=A_b$, and $M_{\ti D}=1.1\,M_{\ti Q}$.
As can be seen, decays into bosons may have large 
branching ratios \cite{sq2dec}.
For the stau decays we choose $\mstau{1}=250$ GeV, $\mstau{2}=500$ GeV,  
and $\mstau{L}<\mstau{R}$. 
Figure~\ref{fig:BRstau}\,a shows the sum of the branching ratios of 
the bosonic $\stau_2$ decays, 
$\sum \mbox{BR}[\stau_2\to \stau_1 + (Z^0,\,h^0,\,H^0,\,A^0), 
                         \:\snu_\tau + (W^-,\,H^-)]$, 
in the $A_\tau$--$\mu$ plane for $\tan\b=30$. 
Figure~\ref{fig:BRstau}\,b shows the $\tan\b$ dependence of the individual 
$\stau_2$ branching ratios for $A_\tau=800$ GeV  and $\mu=1000$ GeV. 
Here ``Gauge/Higgs + X'' refers to the sum of the gauge and Higgs boson modes.
Quite generally, the decays of $\st_2$, $\sb_i$, and $\stau_2$ into Higgs
or gauge bosons can be significant in a large parameter region due to
the Yukawa couplings and mixings $\st$, $\sb$, and $\stau$.

%------------------------------------------------------------------------------
\section{Conclusions} 
%------------------------------------------------------------------------------
We have presented a rather complete phenomenological study of
production and decays of stops, sbottoms, and staus in 
$e^+ e^-$~annihilation with $\sqrt{s} = 500$ -- 800~GeV. We have emphasized
the advantage of using polarized $e^-$ and $e^+$ beams for a better 
determination of the SUSY parameters. We have shown that at high
luminosity (${\cal L} \sim$ 500 fb$^{-1}$) it is absolutely necessary
to include SUSY--QCD and Yukawa coupling correction because
the cross section can be measured at a few percent level.

%------------------------------------------------------------------------------
\section*{Acknowledgments} 
%------------------------------------------------------------------------------
We thank K. Hidaka, T. Kon, and Y. Yamada for a fruitful
collaboration.
We are grateful to W.~Adam for useful discussions. We thank H.~Nowak
and A.~Sopczak for clarifying correspondence.
This work has been supported in part by the 
``Fonds zur F\"orderung der wissenschaftlichen Forschung'' of Austria, 
project no. P13139--PHY.

%------------------------------------------------------------------------------
\section*{References}
%------------------------------------------------------------------------------

\end{document}